\documentclass[showpacs,preprint]{revtex4}
\begin{document}

\title{ Anamolous reduction of magnetic coercivity of graphene oxide and reduced graphene oxide upon cooling}
\author{K. Bagani, A. Bhattacharya, J. Kaur, A. Rai Chowdhury, B. Ghosh, M. Sardar$^1$ and S. Banerjee\footnote{email:sangam.banerjee@saha.ac.in}}
\address{Surface Physics Division, Saha Institute of Nuclear Physics, 1/AF Bidhannagar, Kolkata-700064, India\\
$^1$Material Science Division, Indira Gandhi Center for Atomic Research, Kalpakkam 603 102, India\\
}

\begin{abstract} In this report we present the temperature evolution of magnetic coercivity of graphene oxide (GO) and reduced graphene oxide (RGO). We report an anamolous decrease in coercivity of GO and RGO with decreasing temperature. We could explain this anamolous behavior invoking the inherent presence of ripple in graphene. We observe antiferromagnetic and ferromagnetic behavior at room temperature for GO and RGO respectively, but at low temperatures both shows paramagnetic behavior.

\end{abstract}
\maketitle

\section{INTRODUCTION}
Graphene has drawn a lot of interest not only for their unique electronic properties \cite{Geim} but also for their unusual magnetic properties
\cite{rao,sepioni,Magneitc,Y. Wang}. Magnetic ordering is typically observed in materials with partially filled d or f shell electrons. Though there exists a large number of carbon based organic magnets with p electrons, the T$_{C}$ is usually below 10 K \cite{Tamura}. That is why observation of room temperature ferromagnetism in graphene related materials is so surprising. Many theoretical works already exist to explain unusual magnetic property of graphene related materials. Various reasons have been put forward for the formations of magnetic moments in graphene, such as, existence of zigzag edges \cite{zigzag}, structural defects (vacancies, adatoms, etc.,)  \cite{defect} etc.,. Zigzag edges induce highly degenerate localized electronic states precisely at the Fermi level. These degenerate states are filled with electrons with the same spin in order to minimize the coulomb repulsion energy (an effective Hund’s coupling). Spin magnetic moments can also originate from the dangling bonds localized at the vacancy sites \cite{Choi}.  J. Zhou et. al. \cite{Zhou} have predicted that moments can arise in semi-hydrogenated graphene sheets. L. Xie et. al. \cite{Xie} observed room temperature ferromagnetism in partially hydrogenated epitaxial graphene. W. Li et. al.  \cite{Li} theoretically showed that addition of monovalent and divalent adatom on graphene can also give rise to magnetic moments. This way of generating magnetic moment follows simply from Lieb's theorem \cite{Lieb}, which says that for half-filled bipartitite lattice systems the electrons in sublattice A and B are antiferromagnetically correlated. The ground state magnetic moment for such a system is, M= (N$_{A}$ -N$_{B}$)$\mu$$_{B}$, where N$_{A}$ and N$_{B}$ are the number of A and B sublattice points. Much less is known about the interaction and ordering if any between such local magnetic moments.  Vozmediano et. al. \cite{Vozmediano} studied the interaction between localized moments mediated by RKKY-like interaction but their calculated transition temperature was far below the room temperature. Possibility of strong long range magnetic coupling via itinerant electrons in defective graphene (vacancy or hydrogen chemisorption) was predicted by Yazyev and Helm \cite{Yazyev} assuming metallicity of the system. L. Pisani et. al. \cite{Pisani} studied more realistic situation of a semiconducting ground state of defective graphene sheet and they suggested that long range spin magnetic ordering is possible in graphene. Although there still persists a controversy that the observed magnetism is not its intrinsic property rather magnetism comes from ferromagnetic impurity present in the sample, experimentally ferromagnetism was observed in HOPG only after proton irradiation \cite{Esquinazi} and hence rules out the role of elemental impurity. Y. Wang et. al. \cite{Y. Wang} observed room temperature ferromagnetism of partially reduced graphene oxide. But however, they did not stress on the evolution of magnetic hysteresis as a function of temperature, even though they observed an anomalous behaviour of increase in coercivity as a function of temperature. Magnetic properties of graphene related materials depend upon their microstructure. Structures of Graphene Oxide (GO) and Reduced Graphene Oxide (RGO) varies depending on the synthesis processes \cite{Mermoux} and consequently the magnetic property also changes \cite{Gao}.

Here, we have studied the magnetic property of GO and partially reduced GO from room temperature (300K) down to 10K. We find that both GO and RGO show hysteresis at room temperature and as the temperature is lowered the hysteresis loop diminishes. In this paper we will address this anomalous behaviour of diminishing of hysteresis loop upon lowering the temperature for both GO and RGO and try to understand the difference in the magnetic behaviour between GO and RGO.

\section{EXPERIMENTAL DETAILS}
Graphene Oxide (GO) was synthesized using modified Hummers method \cite{Hummers}. To prepare GO we used graphite powder, conc. H$_2$SO$_4$, NaNO$_3$ and KMnO$_4$. The reaction of H$_2$SO$_4$ with strong oxidizing agents such as NaNO$_3$ and KMnO$_4$ results in increase in the interlayer separation of the carbon layers due to incorporation/intercalation of oxygen atoms. This reaction leads to formation of graphite oxide flakes having single or few layer sheets. The solution was further ultrasonicated and centrifuged. The supernatant was collected and dried in vacuum drying oven at 80$^{0}$C to obtain single layer graphene oxide samples. These GO sheets were reduced by strong reducing agent hydrazine hydrate to form RGO. The resulting suspension is filtered and washed with water and ethanol and dried in an vacuum oven. The samples were characterised by UV-VIS, FTIR, Raman spectroscopy, and SEM. 

Magnetic measurements were carried out using SQUID MPMS XL (Quantum Design). Magnetization data as a function of temperature and magnetic hysteresis data at three different temperatures 10 K, 100 K and 300 K were taken. The magnetic moment values observed are comparable with that in earlier reports \cite{Y. Wang}.

\section{RESULTS AND DISCUSSION}
Fig. 1 shows the FTIR spectra of both GO and RGO. Inset of the fig. 1 shows the UV-VIS spectroscopy data. We observe that the UV-VIS spectra of GO shows a sharp absorption peak near 238 nm which is shifted to 252 nm upon reduction. The shift of the peak position of RGO is due to restoration of sp$^{2}$ carbon network after removing the functionalized groups as suggested in ref. \cite{UV}. FTIR spectroscopy data (fig.1) also indicates the reduction of oxygen-containing groups from graphene oxide sheet. The characteristic bands of GO around 1051 cm$^{-1}$ (C-O), 1223 cm$^{-1}$ (C-O-C), 1613 cm$^{-1}$ (C=O), 1731 cm$^{-1}$ (O=C–OH) and 3440 cm$^{-1}$ (O-H) are considerably decreased after the reduction which is consistent  with the literature \cite{FTIR}. In fig.2 we show the Raman spectra of GO and RGO showing two distinct peaks G and D. Raman spectroscopy is a very important tool to study the average size of the sp$^{2}$ region in sp$^{2}$/sp$^{3}$ combined material. Graphene oxide shows the G peak near 1605 cm$^{-1}$ and D peak near 1370 cm$^{-1}$ which agrees with the previous reports \cite{Raman1}. Due to the reconstruction of the sp$^2$ network at the oxidized region after reduction, the G peak in the RGO shifts to a lower wave number 1595 cm$^{-1}$ \cite{Raman2}.  In fig.3 we show the SEM images of GO and RGO. From the SEM image we can clearly see that GO shows folded layered structure and for the case of RGO, the sheets appear like unfurled structure. Magnetization vs. temperature (in log-log scale) data of GO and RGO are presented in fig. 4. In this study a considerable magnetic moment has been observed in both samples. We observed less moment in RGO than GO which we shall explain below. In the figure (5 $\&$ 6) we plot the magnetic hysteresis curve measured at three different temperatures (10 K, 100 K and 300 K) for GO and RGO respectively. M-T data on a log-log scale can be fitted with a straight line with a slope of -1, from 10K upto about 60-70 K. Almost no coercivity at 10 K supports paramagnetic behaviour at low temperature (see inset of fig. 5), but with increase in temperature, the hysteresis loop opens up i.e., coercivity increases with temperature which is quite unusual. Another interesting feature we observe is that with reduction of GO, the magnetic moment decreases but coercivity increases. GO has more paramagnetic contribution and does not saturate at high fields, while reduced graphene oxide (RGO) shows approach to saturation at high field along with larger coercivity indicating that upon reduction the magnetism in RGO becomes more cooperative in nature. Several theoretical works have tried to explain ferromagnetism in graphene. M. Wang et. al. shows by DFT calculation that oxidized graphene can be ferromagnetic \cite{M. Wang}. FTIR spectra indicate that graphene oxide has considerable amount of oxygen and OH groups in it. Presence of the functionalized groups (epoxide [-O-], hydroxyl [-OH], carbonyl [-C=O] and carboxyl [-COOH]) on the graphene skeleton leads to disruption of sp$^{2}$ bonded carbon network at many places. Many topological defects  are also formed, leading to local crumpling of planar graphene sheets. Deviation from the planarity decreases the efficiency of $\pi$-overlap which reduces the connectivity of the $\pi$ network and causes an increase in $\pi - \pi$$^{*}$ energy gap of the localized double bonds, hence reducing the ring current diamagnetism. Therefore magnetic susceptibility/moment can increase upon oxidation of graphene \cite{Makarova}. On the other hand, disruption of the hexagonal sp$^{2}$ bonded carbon network presumably create many topological defects and isolated dangling bonds which carry local moments \cite{M. Wang} and also functionalized oxygenic groups (-O-H) can introduce localized magnetic moments \cite{Li}.

After on reduction of GO, these functionalized groups are removed. This leads to two effects, (1) the $\pi$ electron with which the functional group was bonded is released as a carrier, and (2) the local crumpling of the planar carbon network caused due the presence of functional group is cured and locally the planar structure is restored. We can assume that there will be lesser oxygenic groups, defects, ruptures in the RGO compared to GO.  We find lesser paramagnetic moments in RGO compared to GO as can be  observed from fig. 4 of M versus T plot.  This gives an indication that the magnetization is proportional to the amount of defects such as oxygen and hydroxyl impurities. Thus, the magnetic moments in GO and RGO is defect in origin. 

Pure graphene is electronically half filled (one $\pi$ electron per site) on a bipartite lattice. The $\pi$ electrons on A and B sublattice are antiferromagnetically correlated. GO and RGO are electronically insulating and so a description in terms of  Hubbard type model in real space rather than extended band electron picture is more appropriate. It has been shown that a strong impurity potentials like  vacancy or any chemisorbed defect induces short range antiferromagnetic (ferrimagnetic) order around itself in a Hubbard model on a half filled honeycomb lattice (just like graphene) \cite{Kumazaki}. The net magnetic moment in a short range antiferromagnetic cluster is given by, M$\propto$$(N_{A}$ -N$_{B})$ as mentioned earlier. Oxygen atoms bonds with two near neighbor C atoms of opposite sublattices and hence do not give rise to any local moment.  OH on the other hand binds with only one C atom (either A or B sublattice) on graphene,  This  can create a local imbalance in N$_A$ and N$_B$ leading to a local moment \cite{Kumazaki}. In a similar way atomic vacancy (missing C atom) can also create magnetic moment. We assume that GO sheets have many patches of OH clusters, This patches are structurally distorted and electrically insulating locally because of gap at the local Fermi level. Nearest neighbour interaction between these $\pi$ electrons within such a patch is antiferromagnetic (superexchange). The magnetic moment of such a small patch is $M=(N_A-N_B) \mu_B $. This uncompensated moment resides at the boundary of the patch. It is like an uncompensated nanosized antiferromagnetic particle with a net moment.  Most of these patches have very small net moments and behave paramagnetic. GO has more paramagnetic moment compared to RGO. This seems natural as reduction process removes many such small moment carrying OH clusters from GO.
After reduction many OH groups are removed and hence we can consider less clustering of OH groups and isolated OH groups are scattered all over the sheet. Since GO has more OH groups than RGO, the net magnetic moment of GO is higher than RGO. At low temperatures these isolated moments randomly placed in an insulating matrix with very little interaction between them due to highly crumpled structure behaves like paramagnets. It has been observed experimentally \cite{ripple} that the crumpling/wrinkling of graphene sheets decreases with increase in temperature. Thus, with increase in temperature, crumpling is reduced and interaction between moments increases and hence deviation from paramagnetic behavior is observed. From fig.4 it can be seen that at low temperature due to increase in crumpling both GO and RGO are paramagnetic, but at higher temperature(above 60-70 K) we see for GO, the moments drops below the paramagnetic slope (see fig.4) and for RGO, the moments increases above the paramagnetic slope. Hence GO and RGO behaves like antiferromagnetic and ferromagnetic/ferrimagnetic respectively.

Now we discuss the M-H loop measurements. At 10 K the M-H curve is paramagnetic for both GO and RGO showing no coercivity. At 100 K and 300 K both the samples shows clear hysteresis loops. We observe that the coercivity increases with increase in temperature which is very unusual. Also the coercivity of RGO is higher than that of GO at these temperatures. As we discussed earlier, both GO and RGO can be considered to have large number of para moments coming from either defects (vacancies) or isolated OH groups. They might also have many clusters of OH groups which can be thought of as nanosized single domain particles. It is these  blocked super moments  which contributes to irreversibility and hysteresis. Here the magnetization curve is known to arise from a competition between the domain anisotropy energy and the Zeeman energy (Stoner Wolfarth model \cite{Stoner}). The coercive field is proportional to $K/M_s$, where $K$ is the anisotropy energy density and $M_s$ is the net moment of any patch. If a patch is structurally very much crumpled then the average anisotropy energy of the patch will be small. Magnetocrystalline anisotropy energy is very sensitive to the local perpendicular (to graphene plane) axis. On a patch having large number of O and OH groups, the variation of this axis from point to point is considerable. The coercive field is dependent on the spatial average (over the whole patch) of this anisotropy energy. Average anisotropy energy density $K$ is smaller in GO than in RGO, because GO is structurally more disordered than RGO. RGO has much less number of O and OH groups and hence much less local crumpling that leads to higher anisotropy energy density K for the patch. So the coercive field in RGO is larger than GO at any temperature as observed. Since with increase in temperature the graphene sheets becomes on the average more planar, the average ‘K’ of any patches of GO and RGO increases, leading to increase in coercivity as temperature increases for both the systems. This we believe to be dominant reason for the increase of coercivity with increase in temperature. So far we have not discussed the interaction between the moments. It may be argued that with increase in temperatures, thermally activated carries could be responsible for interaction between the isolated moments, leading to long range magnetic ordering \cite{Vozmediano}. With increasing temperature, more thermally activated electrons will come into play to increase coercivity along with gradual transition from paramagnetic to more nonlinear shape of M-H. Since the gap arises due to distortion, defects etc. in the sp$^{2}$ sheet, hence lesser gap at the Fermi level of RGO than GO could be the reason for increase of coercivity upon reduction.

\section{CONCLUSION}
To summarize, we synthesized GO and RGO chemically and the observed magnetic property of these two system are different.  The paramagnetic part of the magnetic moment is larger in GO compared to RGO. We argue that the origin of para moments are due to small patches of OH on graphene or other topological defects. Reduction process removes many such OH containing regions as well as anneals some local topological defects. This might be the reason for lower paramagnetic moment in RGO compared to GO. At higher temperatures there is finite hysteresis loops and the  magnetic coercivity anomalously increases with increase in temperatures. We have argued that over and above the small defects and small patches of OH groups which gives paramagnetism, there are also many nanosized OH patches having considerable uncompensated moments.  These are blocked along random directions and are responsible for irreversibility , hysteresis, saturation etc.,.   The coercivity of such blocked supermoments are determined by its average local magnetic anisotropy energy. With increase in temperatures the wrinkles on GO and RGO decreases on the average, this leads to an increase in the local  magnetic anisotropy energy. Hence, an increase in coercivity  with increase in temperature is observed.
Larger  coercivity  in RGO compared to GO at any temperatures is due to less number of functional groups than in GO and hence less wrinkles on it. Our arguments is also supported by  the experimental observation of increase in wrinkle formation in graphene with the decrease in temperature.

\section{ACKNOWLEDGEMENT} We would like to thank Prof. A. Roy of IIT Kharagpur for Raman measurement of our samples. We would like to thank Amity University, Noida for encouraging JK and ARC to work in SINP.

\newpage
\centerline{\bf Figure Captions}
\vspace{0.5 cm}

\noindent{\bf Figure 1}: (color online) FTIR spectroscopy data of Graphene Oxide (red) and Reduced Graphene Oxide (blue). Inset: UV-VIS spectroscopy data.

\noindent{\bf Figure 2}: (color online) Raman spectra of Graphene Oxide (GO) and Reduced Graphene Oxide (RGO).

\noindent{\bf Figure 3}: SEM image of GO and RGO 

\noindent{\bf Figure 4}: (color online) M vs T in log-log scale of Graphene Oxide (open square ) and Reduced Graphene Oxide (open circle). 

\noindent{\bf Figure 5}: (color online) Magnetic Hysteresis of GO at temperature 10K, 100K and 300K. Inset (top left): Magnetic Hysteresis of GO and RGO at 10K. Inset (bottom right) Zoom region of the hysteresis curve.

\noindent{\bf Figure 6}: (color online) Magnetic Hysteresis of RGO at temperature 10K, 100K and 300K. Inset (top left): Magnetic Hysteresis of GO and RGO at 300K. Inset (bottom right) Zoom region of the hysteresis curve.

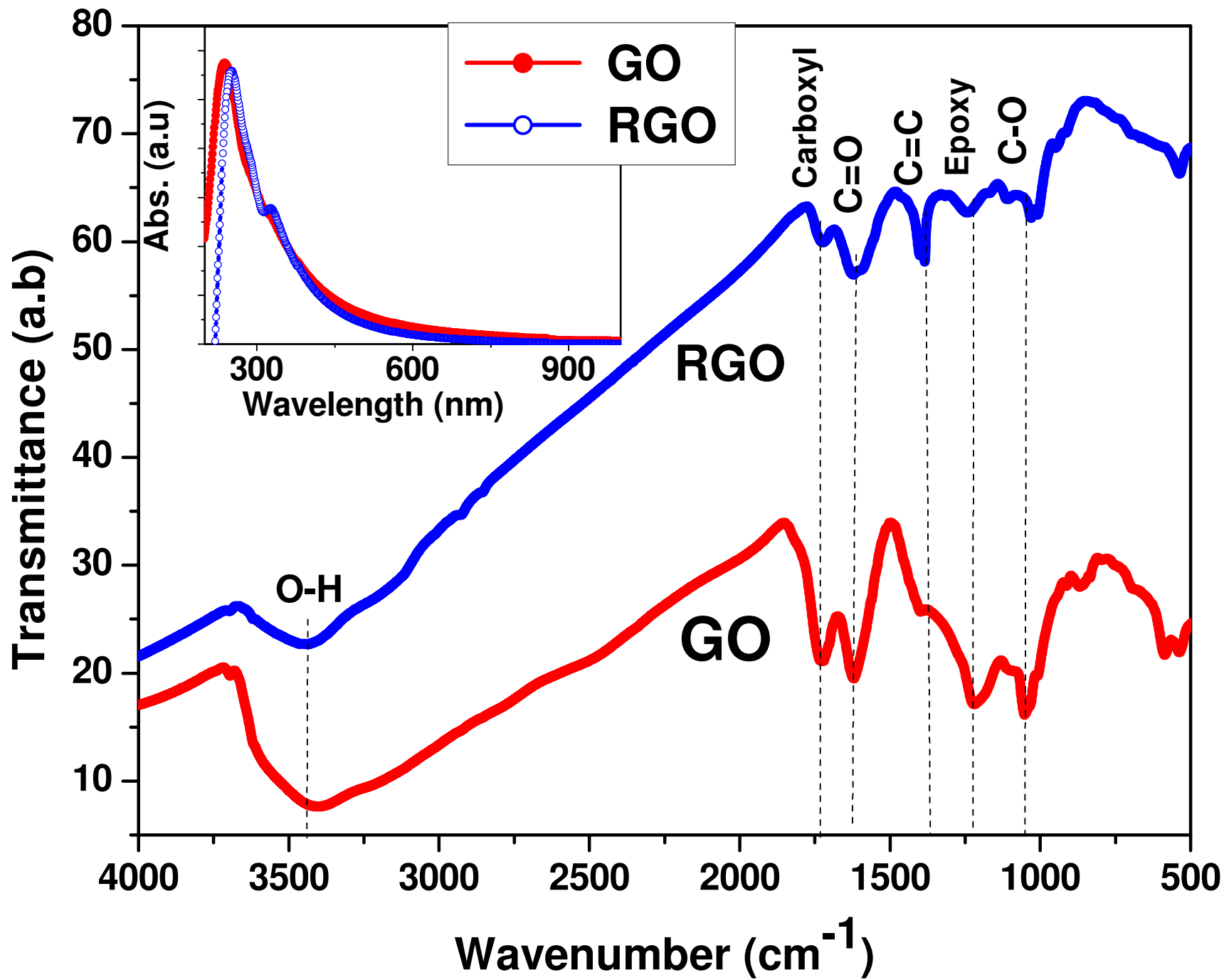
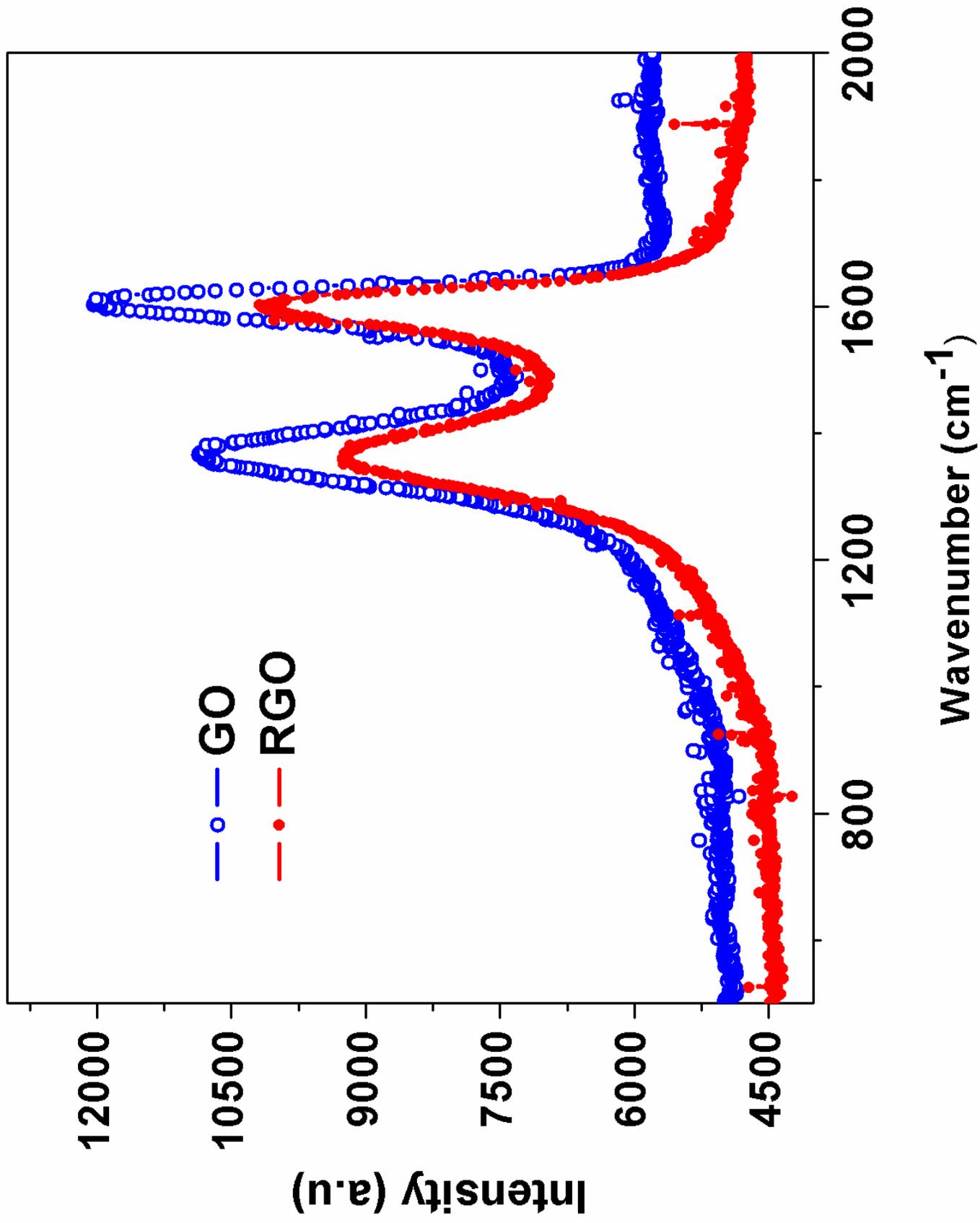
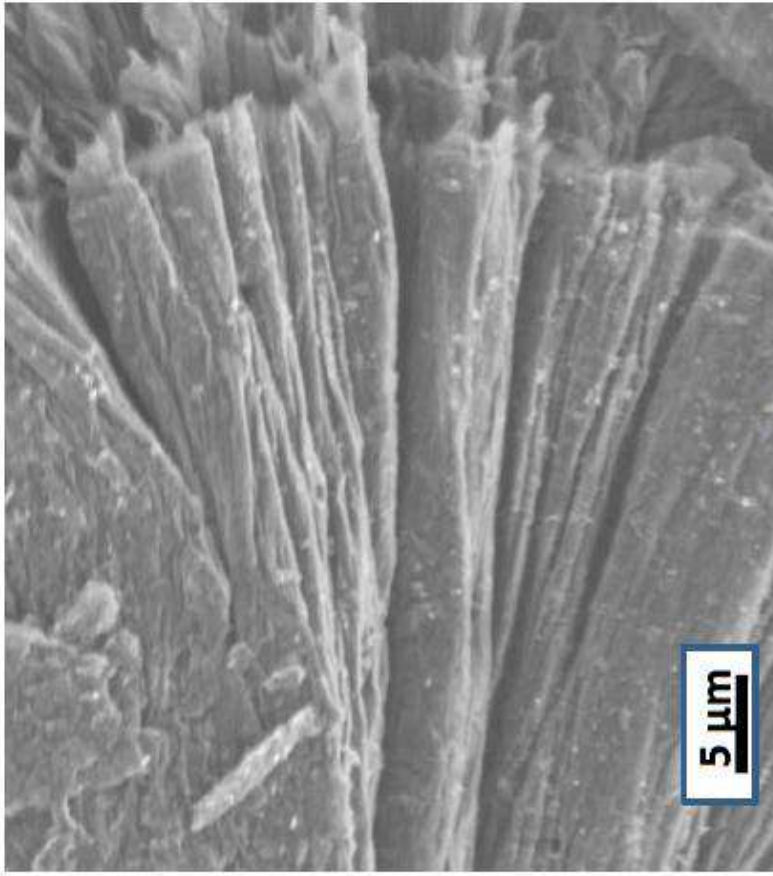
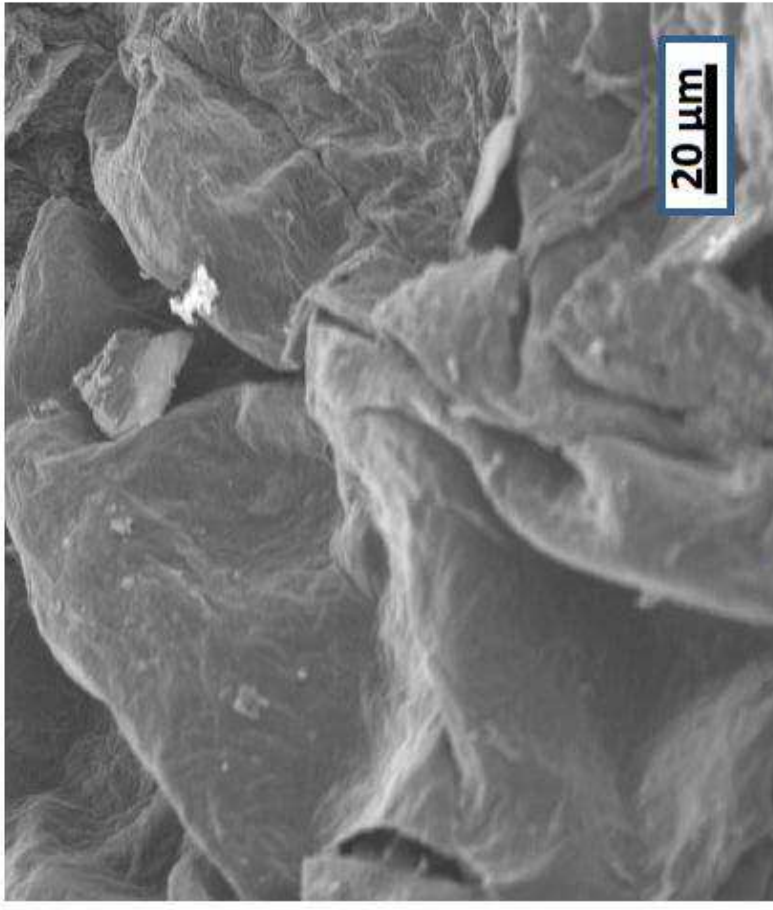
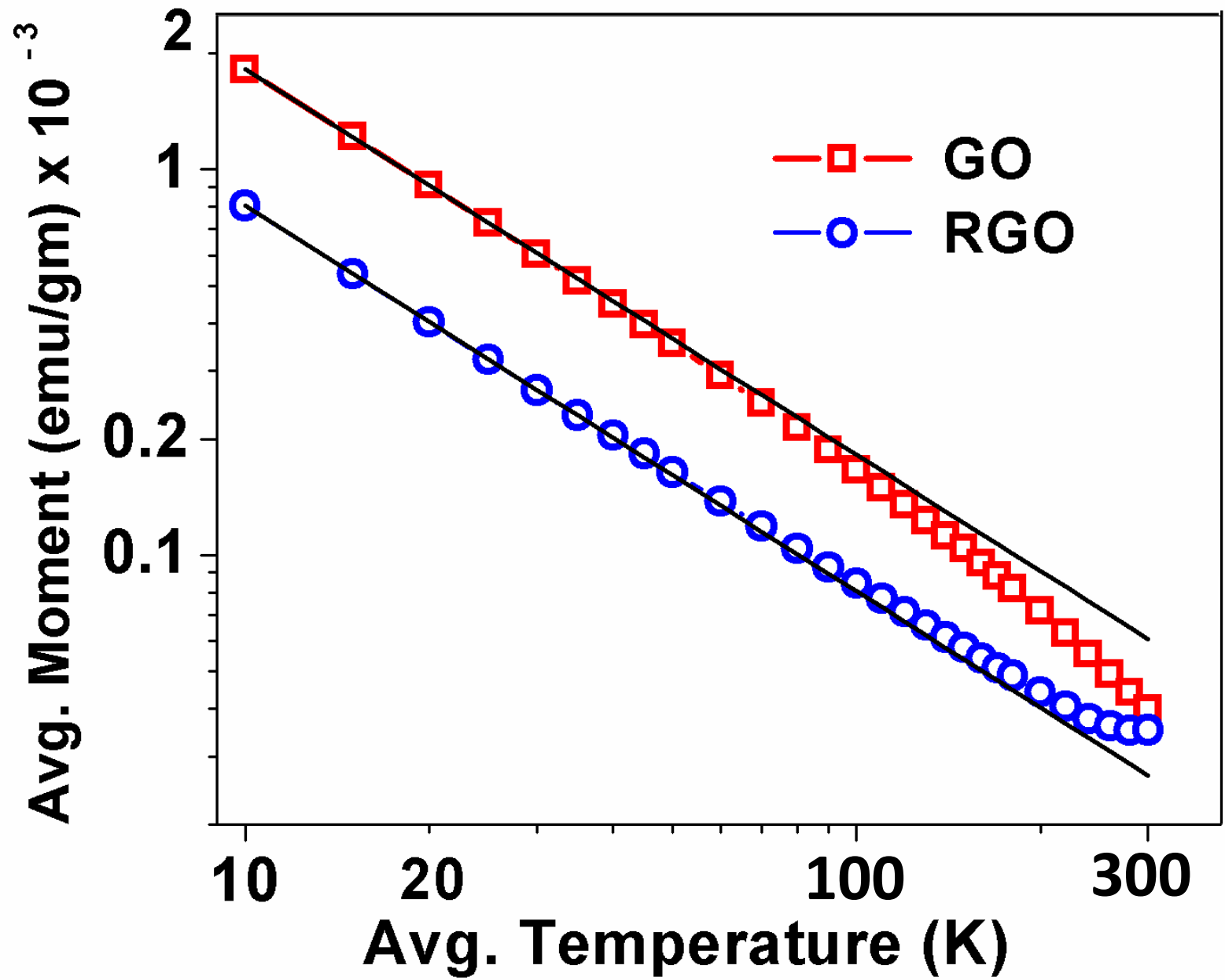
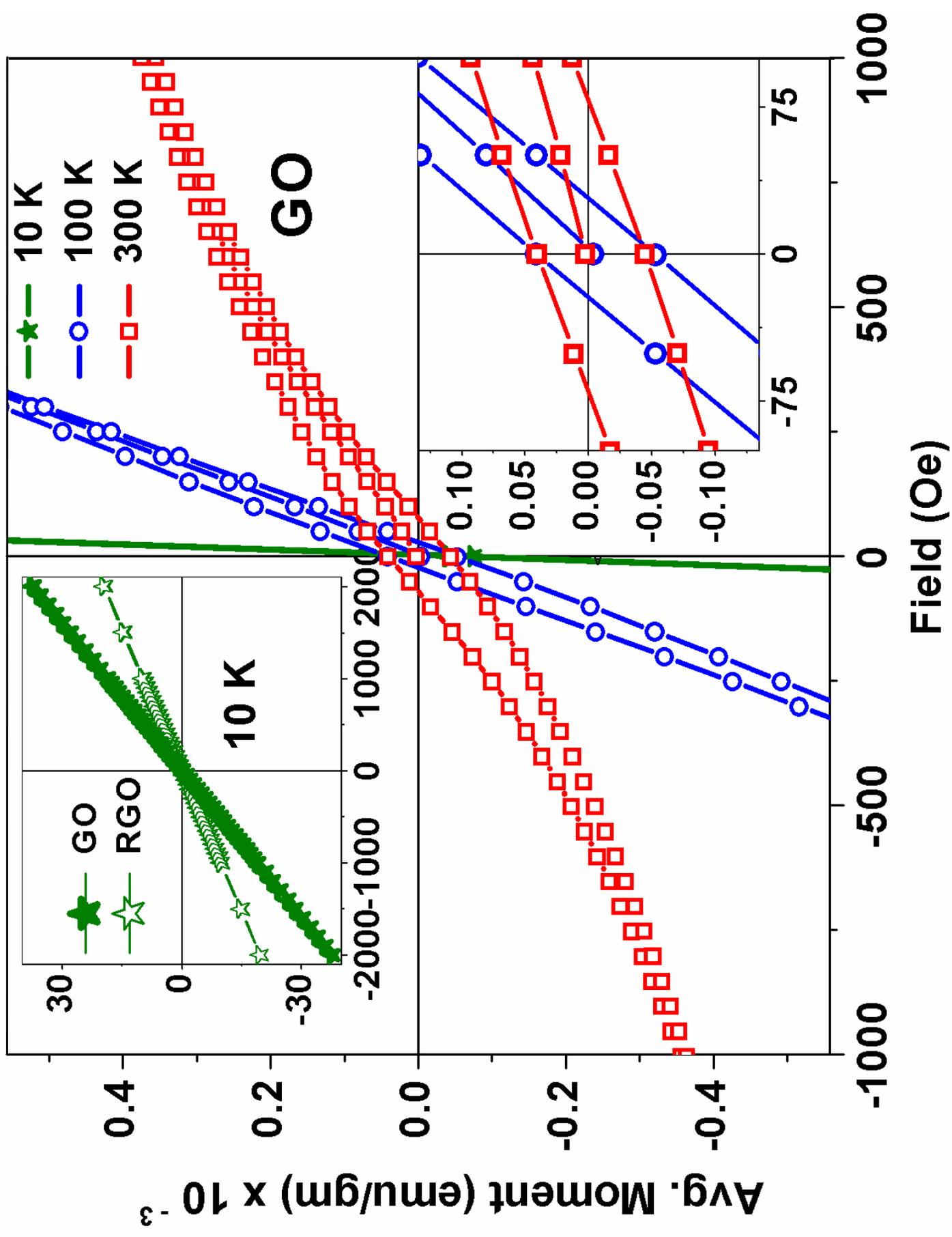
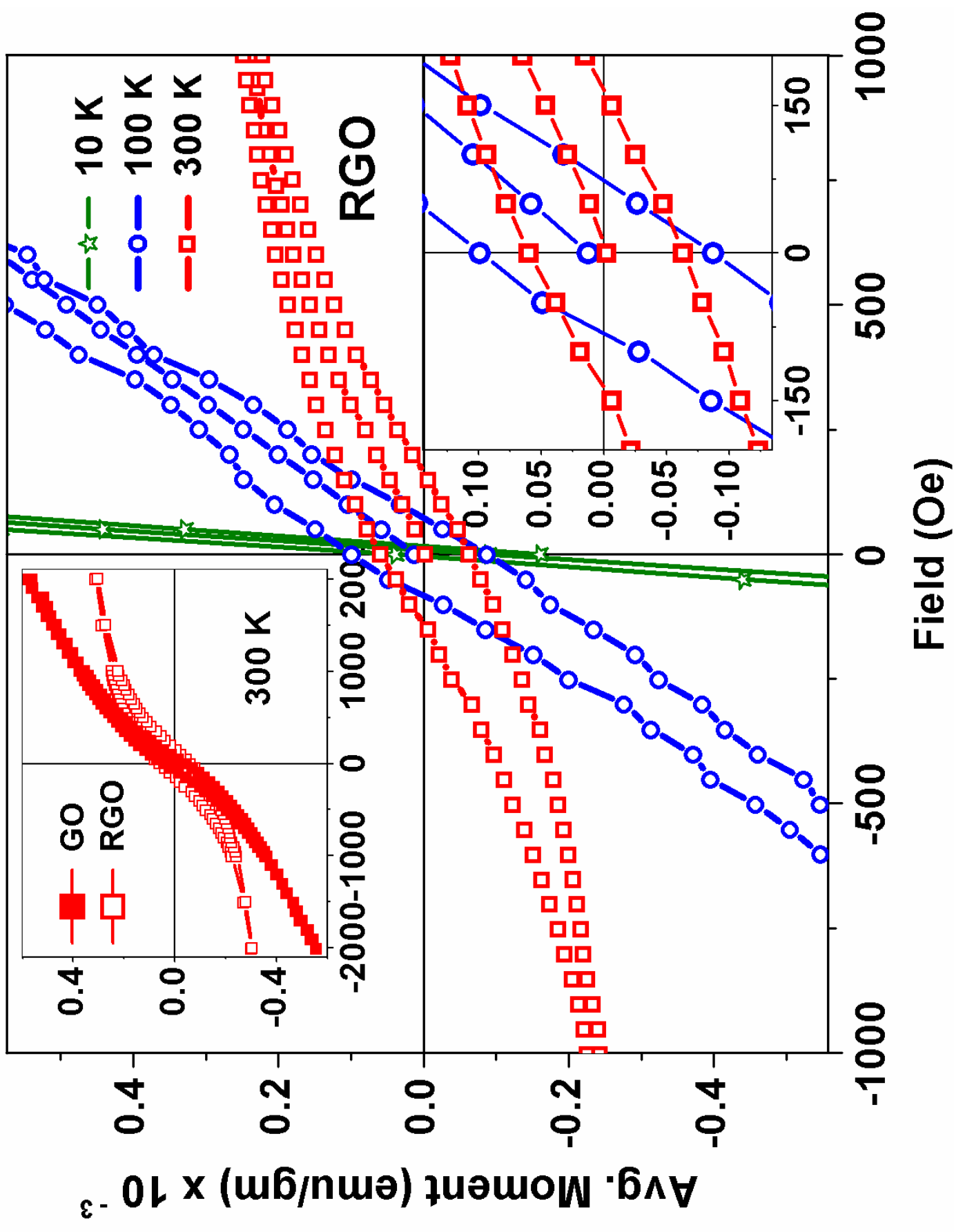
\end{document}